\documentclass[10pt]{extarticle}
\usepackage{graphicx}
\usepackage{amssymb,amsmath}
\def\beq{\begin{eqnarray}}
\def\eeq{\end{eqnarray}}
\usepackage[titletoc,toc,title]{appendix}

\usepackage{amsfonts}

\begin{document}

\title{Exact sum rules for heterogeneous spherical drums}
\author{Paolo Amore \\
\small Facultad de Ciencias, CUICBAS, Universidad de Colima,\\
\small Bernal D\'{i}az del Castillo 340, Colima, Colima, Mexico \\
\small paolo.amore@gmail.com }

\maketitle

\begin{abstract}
We have obtained explicit integral expressions for the sums of inverse powers of the eigenvalues 
of the Laplacian on a  unit sphere, in presence of an arbitrary variable density. 
The exact expressions for the sum rules are obtained by properly "renormalizing" the series, excluding
the divergent contribution of the vanishing lowest eigenvalue.  
For a non--trivial example of a variable density we have applied our formulas to calculate the exact 
sum rules of order two and three, and we have verified these results calculating the sum rules numerically  
using the eigenvalues obtained with the Rayleigh-Ritz method.
\end{abstract}



\section{Introduction}
\label{intro}

In this paper we study the sum rules obtained summing the inverse powers of the eigenvalues
of the Helmholtz equation on a heterogeneous sphere. For the special case of constant density
(to which we can arbitrarily assign the value $\Sigma=1$), one obtains that the eigenfunctions 
are the spherical harmonics, $Y_{lm}(\theta,\phi)$, with  $|m| \leq l$ and $l=0,1,2,\dots$, 
and the corresponding eigenvalues, $l (l+1)$, are $2l+1$ degenerate. 

The sum rule of order $p$ is thus defined in terms of the eigenvalues as~\footnote{Note that to obtain a finite result, 
	the zero mode, corresponding to $l=0$,  has to be excluded from the series.}
\begin{equation}
Z_p = \sum_{l=1}^\infty \frac{2l+1}{(l (l+1))^p}
\end{equation}
with $p>1$.

For the case of an arbitrary density, $\Sigma(\theta,\phi)>0$ over the sphere, however,
the approach outlined above cannot be adopted, since it requires to calculate exactly 
{\sl each} of the eigenvalues of the Helmholtz equation. A similar situation occurrs
in the calculation of the sum rules for quantum billiards on finite domains in the plane,
since the eigenvalues are known exactly only for a limited number of shapes (rectangle, circle, ellipse and 
symmetric circular annulus, among others).
In particular, Itzykson, Moussa and Luck~\cite{Itzykson86} were able to obtain explicit integral
expressions for the sum rules of inverse powers of the Dirichlet eigenvalues of the Laplacian 
on arbitrary domains in two dimensions using a conformal transformation from the domain to the unit
disk, without having to know the eigenvalues exactly. Berry~\cite{Berry86} 
applied the method of Ref.~\cite{Itzykson86} to Aharonov-Bohm quantum billiards, obtaining explicit 
expressions for different shapes. Steiner \cite{Steiner87} also discussed the sum rule for Aharonov-Bohm quantum billiard
of circular shape, extending an approach previously developed in \cite{Steiner85} for confinement potentials.

More recently Kvitsinky has considered the spectral sum rules for nearly circular domains, particularly a $N$-sided regular polygons~\cite{Kvitsinsky96}; Dittmar~\cite{Dittmar02} has obtained the sum rules for fixed and free membrane problems for simply connected domains of the plane, conformally transforming the domain into the unit disk. Sum rules for specific domains are obtained in Ref.~\cite{Dittmar11}. Dostani\'c~\cite{Dostanic11} has obtained the regularized trace of the inverse Dirichlet laplacian on
a bounded convex domain. 

In a series of papers, Refs.~\cite{Amore13A, Amore13B, Amore14, Amore18}, we have derived 
general integral expressions for the spectral sum rules of  inhomogenous strings and membranes, 
for different boundary conditions; the case of Neumann or periodic boundary conditions, discussed
in Ref.~\cite{Amore14}, requires a careful handling of the traces, which are in principle ill defined due to the  singular contribution stemming from the zero mode. Ref.~\cite{Amore18}, finally, introduces a  "regularized" sum rule, which is obtained 
exploting the symmetries of a problem or different boundary conditions. 
The purpose of this paper is to extend the approach outlined in Ref.~\cite{Amore14} to the case of the heterogenous sphere.

The paper is organized as follows: in section \ref{general} we describe the general approach and 
define the sum rules in terms of the appropriate traces; in section \ref{pt} we obtain the
perturbative corrections to the energy of the lowest mode ("zero-mode"); in section \ref{order}
we derive the general integral expressions for the sum rules of order two and three, explicitly proving that
all divergent contributions cancel out, and  we apply these results to a non-trivial example. 
Finally, in section \ref{conclusions}, we draw our conclusions and discuss future work.
The expressions for the perturbative corrections to the energy of the fundamental mode 
and for the integrals appearing in the sum rules of order two and tree for an arbitrary density 
are reported in  the Appendices \ref{appA} and \ref{appB} respectively.

\section{Exact sum rules: general expressions}
\label{general}

Our starting point is the Helmholtz equation on a unit 2-sphere, in presence of a 
variable density 
\begin{equation}
-\Delta \psi_n(\theta,\phi)  = E_n \Sigma(\theta,\phi) \psi_n(\theta,\phi) \label{eq_Helmoltz_dens}
\end{equation} 
where 
\begin{equation}
\Delta \equiv \frac{1}{\sin \theta} \frac{\partial}{\partial \theta} {\sin \theta} \frac{\partial}{\partial \theta} +  \frac{1}{\sin^2 \theta} \frac{\partial^2}{\partial \phi^2} 
\nonumber
\end{equation}
is the angular part of the spherical Laplacian operator.

As discussed in Ref.~\cite{Amore10}, one can define $\Phi_n=\sqrt{\Sigma} \psi_n $ 
and cast this equation into the equivalent form
\begin{equation}
\frac{1}{\sqrt{\Sigma}} (-\Delta) \frac{1}{\sqrt{\Sigma}} \Phi_n(\theta,\phi) = E_n \Phi_n(\theta,\phi)
\label{eq_Helmoltz_dens2}
\end{equation} 
in terms  of the hermitian operator $\hat{O} \equiv \frac{1}{\sqrt{\Sigma}} (-\Delta) \frac{1}{\sqrt{\Sigma}}$. 

Since the lowest eigenvalue of $\hat{O}$ vanishes, it is convenient to introduce 
the modified operator, following Ref.~\cite{Amore14},
\begin{equation}
\hat{O}_\gamma \equiv \frac{1}{\sqrt{\Sigma}} (-\Delta +\gamma) \frac{1}{\sqrt{\Sigma}}
\end{equation}
where $\gamma$ is a constant parameter which will be eventually sent to zero.

Our ultimate goal is to obtain the Green's function associated with $\hat{O}_\gamma$ on the unit sphere; the first
step in this direction is to write the Green's function associated with the operator $(-\Delta+ \gamma)$ on the unit sphere, 
which obeys the  spectral decomposition
\begin{equation}
G_\gamma(\theta,\phi,\theta',\phi')  = \frac{1}{4\pi \gamma} + \sum_{l=1}^\infty \sum_{m=-l}^l 
\frac{Y_{lm}(\theta,\phi)Y_{lm}^\star(\theta',\phi')}{l (l+1)+\gamma}
\end{equation}

In particular, for $\gamma \rightarrow 0$ one can write
\begin{equation}
G_\gamma(\theta,\phi,\theta',\phi')  = \frac{1}{4\pi \gamma} + \sum_{q=0}^\infty (-\gamma)^q 
G^{(q)}(\theta,\phi,\theta',\phi')  
\end{equation}
where
\begin{equation}
G^{(q)}(\theta,\phi,\theta',\phi')  \equiv
\sum_{l=1}^\infty \sum_{m=-l}^l  \frac{Y_{lm}(\theta,\phi)Y_{lm}^\star(\theta',\phi')}{(l (l+1))^{q+1}}
\end{equation}

These functions obey the properties:
\begin{equation}
\begin{split}
-\Delta G^{(0)}(\theta,\phi,\theta',\phi') &= \frac{\delta(\phi-\phi') \delta(\theta-\theta')}{\sin\theta} - \frac{1}{4\pi} \nonumber \\
-\Delta G^{(q)}(\theta,\phi,\theta',\phi') &= G^{(q-1)}_\gamma(\theta,\phi,\theta',\phi') \hspace{1cm} , \hspace{1cm} q=1,2,\dots \nonumber 
\end{split}
\end{equation}
and
\begin{equation}
G^{(q+1)}(\theta,\phi,\theta',\phi')  = 
\int d\Omega''  G^{(0)}(\theta,\phi,\theta'',\phi'') G^{(q)}(\theta'',\phi'',\theta',\phi') \nonumber
\end{equation}

Notice that $G^{(q)}(\theta,\phi,\theta',\phi')$ ($q=0,1,\dots$) are {\sl finite} since they do not contain
contributions from the mode $l=0$ (in Refs.~\cite{Amore13A,Amore13B,Amore14} we actually referred to 
$G^{(0)}(\theta,\phi,\theta',\phi')$ as to a "regularized" Green's function).

Using the property
\begin{equation}
\sum_{m=-l}^l Y_{lm}(\theta,\phi) Y_{lm}^\star(\theta',\phi') = \frac{2 l+1}{4\pi} P_l(x(\theta,\phi,\theta',\phi'))  
\end{equation}
where $x(\theta,\phi,\theta',\phi') \equiv \hat{e}(\theta,\phi) \cdot \hat{e}(\theta',\phi')$ and  $\hat{e}(\theta,\phi) \equiv \sin\theta \cos \phi \ \hat{i} 
+ \sin\theta \sin \phi \ \hat{j} + \cos\theta \ \hat{k}$, we can cast the Green's functions in the form
\begin{equation}
G^{(q)}(\theta,\phi,\theta',\phi')  = \frac{1}{4\pi}
\sum_{l=1}^\infty   \frac{2l+1}{(l (l+1))^{q+1}} P_l(x(\theta,\phi,\theta',\phi')) 
\end{equation}

It is worth noticing that the Green's function $G^{(0)}(\Omega,\Omega')$ is a special case of the generalized
Green's function discussed in Ref.~\cite{Szmytkowski06}:
\begin{equation}
\bar{G}_L(\Omega,\Omega') \equiv \sum_{\begin{array}{c}
l=0 \\
l \neq L \\
\end{array}} \sum_{m=-l}^l \frac{Y_{lm}(\Omega) Y_{lm}^\star (\Omega')}{L (L+1)-l (l+1)}
\end{equation}
and
\begin{equation}
G^{(0)}(\Omega,\Omega') =  - \bar{G}_0(\Omega,\Omega')
\end{equation}

The explicit expression for $G^{(0)}(\Omega,\Omega')$ is well-known and it can be found 
in Refs.~\cite{Szmytkowski06, Freeden80, Englis98,Gustavsson01}
\begin{equation}
G^{(0)}(x)  = \frac{1}{4\pi} \left[\log 2 - 1- \log(1-x) \right] \nonumber \\
\end{equation}

The formulas for the Green's functions of order one and two can be found in Refs.~\cite{Englis98,Gustavsson01} 
\begin{equation}
\begin{split}
G^{(1)}(x)  &= \frac{1}{4\pi} \left[\log\left(\frac{1-x}{1+x}\right) \log\left(\frac{2}{1+x}\right) - \frac{1}{2} \log^2 \left(\frac{2}{1+x}\right)  +  {\rm Li}_2\left(-\frac{1-x}{1+x}\right) +1 \right] \nonumber \\
G^{(2)}(x)  &= \frac{1}{4\pi} \left[ \frac{\pi^2}{6} -2 +2 \zeta(3) + \log\left(\frac{1-x}{2}\right) {\rm Li}_2\left(\frac{1-x}{2}\right)  - {\rm Li}_2\left(\frac{1+x}{2}\right) - 2 {\rm Li}_3\left(\frac{1-x}{2}\right) \right] \nonumber  
\end{split}
\end{equation}
where ${\rm Li}_\nu(z) \equiv \frac{z}{\Gamma(\nu )}\int_0^\infty \frac{t^{\nu -1}}{\left(e^t-z\right)} dt$
is the polylogarithm of order $\nu$ ($\nu >0$).

The Green's function associated with $\hat{O}_\gamma$ can be now expressed  as
\begin{equation}
G_{\hat{O}_\gamma}(\theta,\phi,\theta',\phi') = \sqrt{\Sigma (\theta,\phi)}  G_\gamma (\theta,\phi,\theta',\phi')  \sqrt{\Sigma (\theta',\phi')}
\end{equation}
since
\begin{equation}
\begin{split}
\hat{O}_\gamma G_{\hat{O}_\gamma} &= \frac{1}{\sqrt{\Sigma (\theta,\phi)}}  (-\Delta+\gamma) G_\gamma (\theta,\phi,\theta',\phi')  \sqrt{\Sigma (\theta',\phi')} \nonumber \\
=& \frac{1}{\sqrt{\Sigma (\theta,\phi)}} \frac{\delta(\theta-\theta') \delta(\phi-\phi')}{\sin\theta}  \sqrt{\Sigma (\theta',\phi')}  \nonumber \\
=&  \delta (\Omega-\Omega')
\end{split}
\end{equation}

Exploiting the invariance of the trace with respect to unitary transformations and using the completeness of the basis of the homogeneous
problem we can write the sum rule~\cite{Amore13A,Amore13B,Amore14} 
\begin{equation}
Z_p(\gamma) \equiv \sum_{n=0}^\infty \frac{1}{E_n(\gamma)^p}
\end{equation}
as
\begin{equation}
Z_p(\gamma) = \int \ G_{\hat{O}_\gamma}(\Omega_1,\Omega_2) \ \dots \  G_{\hat{O}_\gamma}(\Omega_p,\Omega_1) \  d\Omega_1 \dots d\Omega_p 
\label{eq_trace}
\end{equation}

Unfortunately, eq.~(\ref{eq_trace}) is not very useful since it diverges as $\gamma \rightarrow 0$, due to
the singular behavior of $G_{\hat{O}_\gamma}$ in this limit. For this reason 
it is then convenient to introduce the regularized sum rule 
\begin{equation}
\tilde{Z}_p(\gamma) = \sum_{n=1}^\infty \frac{1}{E_n^p(\gamma)} = Z_p(\gamma) - \frac{1}{E_0(\gamma)^p}
\label{eq_Zp}
\end{equation}
by taking out the contributions stemming from the zero mode (we will discuss soon the calculation of 
$E_0(\gamma)$ using perturbation theory for $|\gamma| \ll 1$). 

Since $\tilde{Z}_p(\gamma)$ is now well behaved for $\gamma \rightarrow 0$, we conclude that
\begin{equation}
\sum_{n=1}^\infty \frac{1}{E_n^p} = \lim_{\gamma \rightarrow 0} \left[ Z_p(\gamma) - \frac{1}{E_0(\gamma)^p}\right]
\end{equation}

Some remarks:
\begin{itemize}
\item For $\gamma \rightarrow 0$, $Z_p(\gamma)$ and $1/E_0(\gamma)^p$ can be Laurent expanded  around $\gamma=0$:
\begin{equation}
\begin{split}
Z_p(\gamma) &= z_{-p} \gamma^{-p} +z_{-p+1} \gamma^{-p+1} + \dots + z_0 + z_1 \gamma + \dots \nonumber \\
\frac{1}{E_0(\gamma)^p} &= \epsilon_{-p} \gamma^{-p} +\epsilon_{-p+1} \gamma^{-p+1} + \dots + \epsilon_0 + \epsilon_1 \gamma + \dots \nonumber
\end{split}
\end{equation}
\item The finiteness of $\lim_{\gamma\rightarrow 0} \tilde{Z}_p(\gamma)$ requires that
\begin{equation}
z_{-p} = \epsilon_{-p} \hspace{.5cm} , \hspace{0.5cm} z_{-p+1} = \epsilon_{-p+1}
 \hspace{.5cm} , \hspace{0.5cm}  \dots  \hspace{.5cm} , \hspace{0.5cm}  z_{-1} = \epsilon_{-1} \nonumber
\end{equation}
\item  The singular (for $\gamma \rightarrow 0$) part of the heterogeneous Green's functions appearing in eq.~(\ref{eq_Zp})   may contribute to $z_0$ as long as it combines with suitable contributions from the remaining Green's functions, that are vanishing 
with the appropriate strength (of course this is not the case if the spectrum does not contain a zero mode). It is easy
to check that $Z_p(\gamma)$ contains at most Green's functions of order $p+1$, $G^{(p+1)}$;

\item The calculation of $\tilde{Z}_p$ requires calculating the lowest eigenvalue using perturbation theory up to order $p+1$:
\begin{equation}
E_0 = E_0^{(1)} \gamma + E_0^{(2)} \gamma^2 + \dots \nonumber
\end{equation}
from which
\begin{equation}
\begin{split}
\frac{1}{E_0(\gamma)^p} &= \frac{1}{\gamma^p (E_0^{(1)})^p} - p \frac{E_0^{(2)}}{\gamma^{p-1} (E_0^{(1)})^{p+1} }  \nonumber \\ 
+& 
\frac{p}{2}\frac{1}{\gamma^{p-2} (E_0^{(1)})^{p+2}}
\left((p+1) (E_0^{(2)})^2-2 E_0^{(1)} E_0^{(3)}\right) + \dots \nonumber
\end{split}
\end{equation}

Specifically, for $p=2$ and $p=3$ one has
\begin{equation}
\begin{split}
\frac{1}{E_0(\gamma)^2} =& \frac{1}{\gamma ^2 \left[E_0^{(1)}\right]^2}-\frac{2 E_0^{(2)}}{\gamma  \left[E_0^{(1)}\right]^3} +\frac{3 \left[ E_0^{(2)}\right]^2-2 E_0^{(1)} E_0^{(3)}}{\left[E_0^{(1)}\right]^4} + O(\gamma) \nonumber \\
\frac{1}{E_0(\gamma)^3} =&  \frac{1}{\gamma ^3 \left[E_0^{(1)}\right]^3}
-\frac{3 E_0^{(2)}}{\gamma ^2 \left[E_0^{(1)}\right]^4}
-\frac{3 \left(-2 \left[ E_0^{(2)}\right]^2+E_0^{(1)} E_0^{(3)}\right)}{\gamma \left[E_0^{(1)}\right]^5}\nonumber \\
+& \frac{-10 \left[ E_0^{(2)}\right]^3+12 E_0^{(1)} E_0^{(2)} E_0^{(3)}-3 \left[E_0^{(1)}\right]^2 E_0^{(4)}}{\left[ E_0^{(1)} \right]^6} + O(\gamma) \nonumber
\end{split}
\end{equation}

\item The explicit expressions for the exact sum rules of order $2$ and $3$ are
\begin{equation}
\begin{split}
\tilde{Z}_2 &= \int G^{(0)}(\Omega_1,\Omega_2) \Sigma(\Omega_2)  G^{(0)}(\Omega_2,\Omega_1) \Sigma(\Omega_1) d\Omega_1 d\Omega_2 \nonumber \\
-& \frac{1}{2\pi} \int \Sigma(\Omega_1) G^{(1)}(\Omega_1,\Omega_2) \Sigma(\Omega_2) d\Omega_1 d\Omega_2
\nonumber \\
-& \frac{3 \left[ E_0^{(2)}\right]^2-2 E_0^{(1)} E_0^{(3)}}{\left[E_0^{(1)}\right]^4}  \nonumber \\
\tilde{Z}_3 =& \int G^{(0)}(\Omega_1,\Omega_2) \Sigma(\Omega_2)  G^{(0)}(\Omega_2,\Omega_3) \Sigma(\Omega_3) G^{(0)}(\Omega_3,\Omega_1) \Sigma(\Omega_1)  d\Omega_1 d\Omega_2 d\Omega_3 \nonumber \\
-& \frac{3}{2\pi} \int \Sigma(\Omega_1)  G^{(1)}(\Omega_1,\Omega_2) \Sigma(\Omega_2)   G^{(0)}(\Omega_2,\Omega_3)  \Sigma(\Omega_3)   d\Omega_1 d\Omega_2 d\Omega_3
\nonumber \\
+& \frac{3}{(4\pi)^2} \left( \int \Sigma(\Omega_1)  G^{(2)}(\Omega_1,\Omega_2)  \Sigma(\Omega_2)  d\Omega_1 d\Omega_2 \right) \ \left( \int \Sigma(\Omega_3) d\Omega_3 \right)
\nonumber \\
-& \frac{-10 \left[ E_0^{(2)}\right]^3+12 E_0^{(1)} E_0^{(2)} E_0^{(3)}-3 \left[E_0^{(1)}\right]^2 E_0^{(4)}}{\left[ E_0^{(1)} \right]^6}  \nonumber 
\end{split}
\end{equation}

The expressions for higher order sum rules can be worked out in a completely similar way.

Before being able to cast $\tilde{Z}_2$ and $\tilde{Z}_3$ into a simpler form, we need to apply perturbation theory to derive
the explicit expression for $E_0(\gamma)$ up to a given order. This is done in the next section.

\end{itemize}
 
\section{Perturbation theory for the zero mode} 
\label{pt}

Consider the eigenvalue equation for the lowest mode
\begin{equation}
(-\Delta + \gamma) \psi_0(\Omega) = E_0 \Sigma(\Omega) \psi_0(\Omega)
\label{Helmoltz_eq}
\end{equation}
and assume $\gamma \rightarrow 0$ and
\begin{subequations}
\begin{align}
\label{eqPT1} 
E_0 &=  \sum_{k=1}^\infty E_0^{(k)} \gamma^k  \\
\label{eqPT2} 
\psi_0(\Omega) =& Y_{00}(\Omega) + \sum_{k=1}^\infty \psi_0^{(k)}(\Omega) \gamma^k 
\end{align}
\end{subequations}


By inserting these expressions inside the Helmholtz equation (\ref{Helmoltz_eq})
one obtains a system of equations, one for each order in $\gamma$. 

Starting to zero order, one has the equation
\begin{equation}
-\Delta \psi_0^{(0)} = 0 \label{eq_PT0}
\end{equation}
from which we obtain the leading contributions to the eigenvalue and to the wave function (normalized over
the total solid angle)
\begin{subequations}
\begin{align}
E_0^{(0)} &= 0 \\
\psi_0^{(0)}(\Omega) &= Y_{00}(\Omega)
\end{align}
\end{subequations}

To first order one needs  to solve the equation
\begin{equation}
-\Delta \psi_0^{(1)} + \psi_0^{(0)} = E_0^{(1)} \Sigma \psi_0^{(0)} \label{eq_PT1}
\end{equation}

Using eq.(\ref{eq_PT0}), we can project equation (\ref{eq_PT1}) over the  zero mode, obtaining 
\begin{equation}
E_0^{(1)} = \frac{4\pi}{\int \Sigma(\theta,\phi) d\Omega} \ .
\end{equation}

We now write the first order correction to  the wave function as
\begin{equation}
\psi_0^{(1)}(\theta,\phi) = \sum_{l=1}^\infty \sum_{m=-l}^l c_{lm}^{(1)} Y_{lm}(\theta,\phi)
\end{equation}
and substitute inside  equation (\ref{eq_PT1}). 

With straightforward algebra we obtain
\begin{equation}
\psi_0^{(1)}(\theta,\phi) = \frac{E_0^{(1)}}{\sqrt{4\pi}} \int G^{(0)}(\Omega,\Omega') \Sigma(\Omega') d\Omega'
\end{equation}
where $G^{(0)}$ is the regularized Green's function introduced earlier.

To order $k$ ($k \geq 2$) one obtains the equation
\begin{equation}
(-\Delta )\psi_0^{(k)}(\theta,\phi) + \psi_0^{(k-1)}(\theta,\phi) = \Sigma(\theta,\phi) \sum_{j=1}^{k-1} E_0^{(j)} \psi_0^{(k-j)}(\theta,\phi)
\end{equation}

The corrections of order $k$ to the eigenvalue and to the eigenfunction are obtained as done to order $1$ and they read
\begin{subequations}
\begin{align}
E_0^{(k)} &=  - \frac{\sum_{j=1}^{k-1} E_0^{(j)} \langle \psi_0^{(0)} | \Sigma | \psi_0^{(k-j)} \rangle}{\langle \psi_0^{(0)} |  \Sigma | \psi_0^{(0)} \rangle} \label{eq_EN_recur}\\
\psi_0^{(k)}(\Omega) &=  \sum_{j=1}^k E_0^{(j)} \int d\Omega' G^{(0)}(\Omega,\Omega') \Sigma(\Omega') \psi_0^{(k-j)}(\Omega') \nonumber \\
-& \int d\Omega'  G^{(0)}(\Omega,\Omega') \psi_0^{(k-1)}(\Omega') \label{eq_WF_recur}
\end{align}
\end{subequations}
 
The expressions for the perturbative corrections to the energy of the zero mode
up to fourth order, obtained solving recursively  eqs.~(\ref{eq_EN_recur})
and (\ref{eq_WF_recur}), are reported in  Appendix \ref{appA}.

\section{Exact sum rules of given order}
\label{order}

The final expressions for the sum rules can now be worked out, using the explicit expressions for the perturbative corrections to 
$E_0$ obtained in the previous section. We will concentrate only on the sum rules of order two and three, although similar expressions
can be obtained also for sum rules of higher order.

In particular, for $\gamma \rightarrow 0$, we find that
\begin{equation}
\begin{split}
Z_2(\gamma) &\approx \left(\frac{\int \Sigma(\Omega)d\Omega}{4\pi} \right)^2 \frac{1}{\gamma^2}
+ \frac{\int \Sigma(\Omega) G^{(0)}(\Omega,\Omega') \Sigma(\Omega') d\Omega d\Omega'}{2\pi} \frac{1}{\gamma} \nonumber \\
+& \left[ \int  G^{(0)}(\Omega,\Omega') \Sigma(\Omega') G^{(0)}(\Omega',\Omega) \Sigma(\Omega) d\Omega d\Omega' \right. \nonumber \\
-& \left. \frac{ \int  \Sigma(\Omega) G^{(1)}(\Omega,\Omega') \Sigma(\Omega') d\Omega d\Omega' }{2\pi} \right] + O(\gamma)
\end{split}
\end{equation}
and
\begin{equation}
\begin{split}
\frac{1}{E_0^2(\gamma)} &\approx \left(\frac{\int \Sigma(\Omega)d\Omega}{4\pi} \right)^2 \frac{1}{\gamma^2}
+ \frac{\int \Sigma(\Omega) G^{(0)}(\Omega,\Omega') \Sigma(\Omega') d\Omega d\Omega'}{2\pi} \frac{1}{\gamma} \nonumber \\
+ & \left[ 2 \frac{\int   \Sigma(\Omega)  G^{(0)}(\Omega,\Omega') \Sigma(\Omega') G^{(0)}(\Omega',\Omega'')  \Sigma(\Omega'') d\Omega d\Omega' d\Omega''}{\int \Sigma(\Omega) d\Omega} \nonumber \right. \\
- & \left. \left(\frac{ \int  \Sigma(\Omega) G^{(0)}(\Omega,\Omega') \Sigma(\Omega') d\Omega d\Omega' }{\int \Sigma(\Omega) d\Omega}\right)^2 - \frac{ \int  \Sigma(\Omega) G^{(1)}(\Omega,\Omega') \Sigma(\Omega') d\Omega d\Omega' }{2\pi} \right] \nonumber \\
+ & O(\gamma)
\end{split}
\end{equation}

As a result, we see that the singularities in $\tilde{Z}_2(\gamma)$  cancel identically for $\gamma \rightarrow 0$, as anticipated,  and
the sum rule of order two is therefore 
\begin{equation}
\begin{split}
\sum_{n=1}^\infty \frac{1}{E_n^2} &= 
\int G^{(0)}(\Omega,\Omega')\Sigma(\Omega') G^{(0)}(\Omega',\Omega)\Sigma(\Omega) d\Omega d\Omega' \nonumber \\
-& 2 \ \frac{\int \Sigma(\Omega) G^{(0)}(\Omega,\Omega') \Sigma(\Omega') G^{(0)}(\Omega',\Omega'') \Sigma(\Omega'') 
d\Omega d\Omega' d\Omega''}{\int \Sigma(\Omega) d\Omega}\nonumber \\
+& \left( \frac{\int \Sigma(\Omega) G^{(0)}(\Omega,\Omega')\Sigma(\Omega') d\Omega d\Omega'}{\int \Sigma(\Omega) d\Omega} \right)^2
\end{split}
\end{equation}

In the case of the sum rule of order three we also obtain that the singularities for $\gamma \rightarrow 0$ cancel out
identically inside $\tilde{Z}_3(\gamma)$ and the sum rule reads
\begin{equation}
\begin{split}
\sum_{n=1}^\infty \frac{1}{E_n^3} &= 
\int G^{(0)}(\Omega,\Omega')\Sigma(\Omega') G^{(0)}(\Omega',\Omega)\Sigma(\Omega) G^{(0)}(\Omega',\Omega'')\Sigma(\Omega'') 
d\Omega d\Omega' d\Omega''\nonumber \\
- & 3 \ \frac{\int \Sigma(\Omega) G^{(0)}(\Omega,\Omega') \Sigma(\Omega') G^{(0)}(\Omega',\Omega'') \Sigma(\Omega'')
G^{(0)}(\Omega'',\Omega''') \Sigma(\Omega''')  d\Omega d\Omega' d\Omega'' d\Omega'''}{\int \Sigma(\Omega) d\Omega}\nonumber \\
+ & 3 \ \frac{\left(\int \Sigma(\Omega) G^{(0)}(\Omega,\Omega')\Sigma(\Omega') d\Omega d\Omega' \right)}{\left(\int \Sigma(\Omega) d\Omega\right)^2} \nonumber \\
\cdot &  \left(\int \Sigma(\Omega) G^{(0)}(\Omega,\Omega')\Sigma(\Omega') G^{(0)}(\Omega',\Omega'')\Sigma(\Omega'') d\Omega d\Omega' d\Omega''\right) \nonumber \\
-& \left( \frac{\int \Sigma(\Omega) G^{(0)}(\Omega,\Omega')\Sigma(\Omega') d\Omega d\Omega'}{\int \Sigma(\Omega) d\Omega} \right)^3 
\end{split}
\end{equation}

Using the definitions in Appendix \ref{appB} one has
\begin{equation}
\begin{split}
\sum_{n=1}^\infty \frac{1}{E_n^2} &=  \mathcal{J}_1^{(0,0)} - \frac{1}{2\pi} \mathcal{I}_2^{(0,0)}  + \left(\frac{ \mathcal{I}_1^{(0)} }{4\pi}\right)^2 \\
\sum_{n=1}^\infty \frac{1}{E_n^3} &= \mathcal{J}_2^{(0,0,0)} - \frac{3}{4\pi} \mathcal{I}_3^{(0,0,0)} + \frac{3}{16 \pi^2} \mathcal{I}_1^{(0)} \mathcal{I}_2^{(0,0)} -  \left(\frac{ \mathcal{I}_1^{(0)} }{4\pi}\right)^3 \label{exactsr}
\end{split}
\end{equation}

As an application we consider the density
\begin{equation}
\Sigma(\theta,\phi) = 1 + \kappa \ Y_{10}(\theta,\phi) = 1 + \frac{1}{2} \sqrt{\frac{3}{\pi }} \kappa \cos (\theta )
\nonumber 
\end{equation}
where the requirement $\Sigma(\Omega)>0$ on the sphere implies the condition $|\kappa| < 2 \sqrt{\pi/3} \approx 2.04665$.

We have calculated explicitly the integrals appearing in the sum rules of order two and three:
\begin{equation}
\begin{split}
\mathcal{I}_1^{(0)}     &=  \frac{\kappa^2}{2}  \nonumber \\
\mathcal{I}_2^{(0,0)}   &=  \frac{\kappa^2}{4}  \nonumber \\
\mathcal{I}_3^{(0,0,0)} &=  \frac{\kappa^2}{8} + \frac{\kappa^4}{120 \pi} \nonumber \\
\mathcal{J}_1^{(0,0)}   &= 1 + \frac{\kappa^2}{8 \pi} \nonumber \\
\mathcal{J}_2^{(0,0,0)} &= 2 ( \zeta(3)-1) + \frac{3 \kappa^2}{32 \pi} \nonumber
\end{split}
\end{equation}

The explicit expressions for the sum rules (\ref{exactsr}) are
\begin{equation}
\begin{split}
\sum_{n=1}^\infty \frac{1}{E_n^2} &= 1 + \frac{\kappa^4}{64 \pi^2}  \\
\sum_{n=1}^\infty \frac{1}{E_n^3} &= 2 (\zeta (3)-1) +\frac{11 \kappa^4}{640 \pi ^2}-\frac{\kappa ^6}{512 \pi ^3}
\end{split}
\label{sumruleexact}
\end{equation}

\begin{figure}
\begin{center}
\bigskip\bigskip\bigskip
\includegraphics[width=12cm]{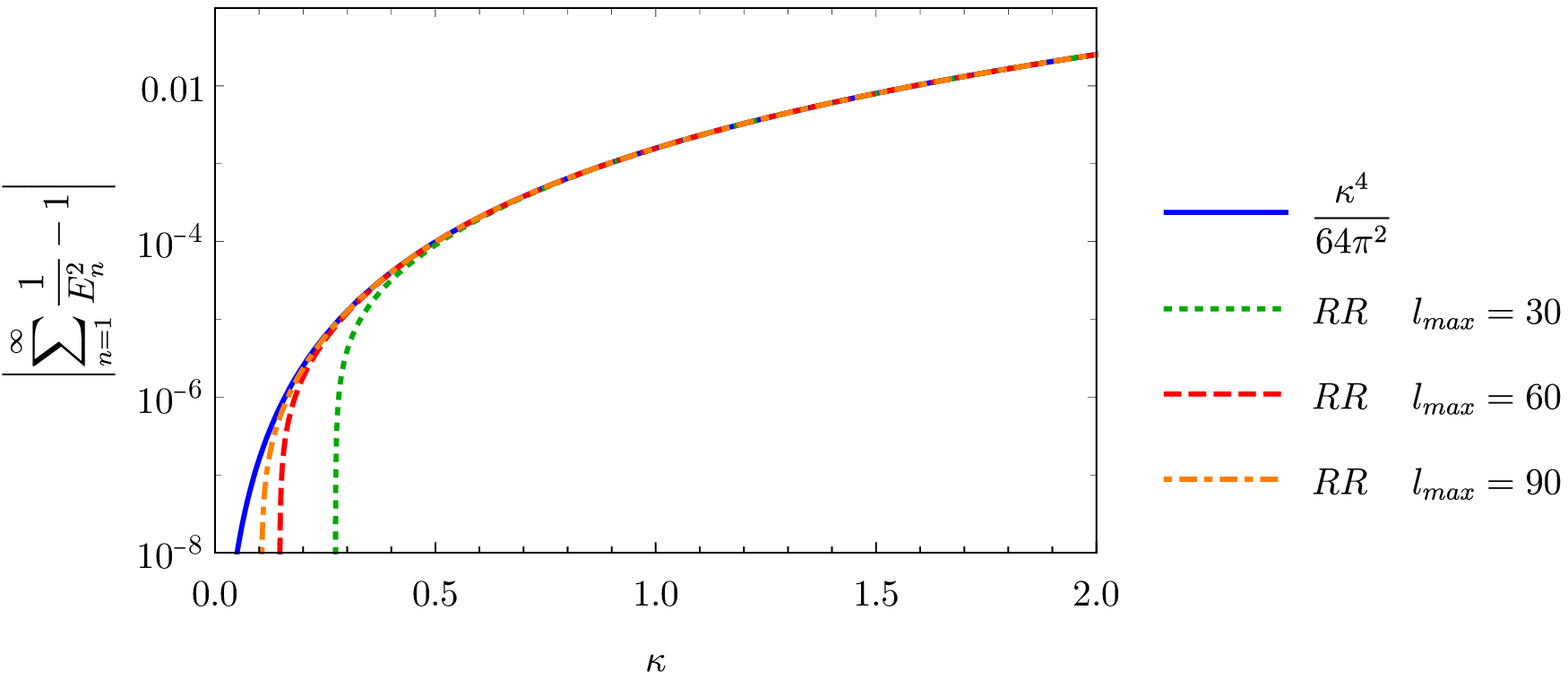} \ \ \  \ 
\caption{$\left| \sum_{n=1}^\infty \frac{1}{E_n^2} - 1 \right|$ as a function of $\kappa$. The solid line is the exact result $\frac{\kappa^4}{64\pi^2}$, while the dotted, dashed and dot-dashed lines are the numerical results obtained approximating the  eigenvalues with the Rayleigh-Ritz method with $l_{max}=30$,$60$ and $90$ respectively.}
\label{Fig_1}
\end{center}
\end{figure}

\begin{figure}
	\begin{center}
		\bigskip\bigskip\bigskip
		\includegraphics[width=12cm]{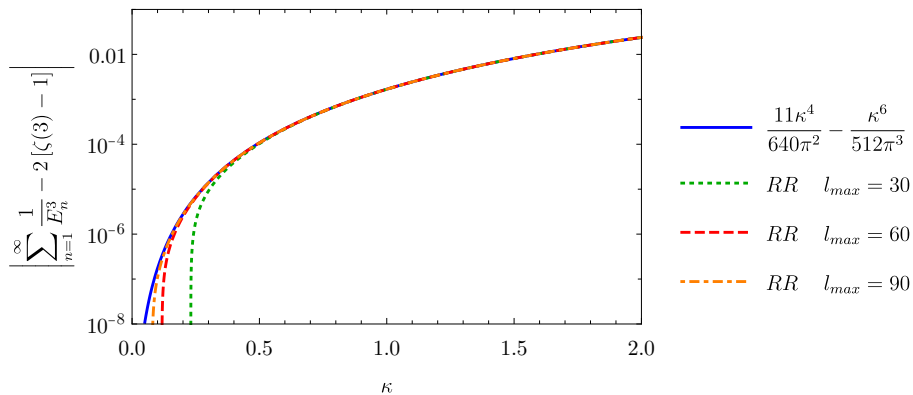}  \ \ \ \
		\caption{$\left| \sum_{n=1}^\infty \frac{1}{E_n^3} - 2 (\zeta(3)-1) \right|$ as a function of $\kappa$. The solid line is the exact result $\frac{\kappa^4}{64\pi^2}$, while the dotted, dashed and dot-dashed lines are the numerical results obtained approximating the  eigenvalues with the Rayleigh-Ritz method with $l_{max}=30$,$60$ and $90$ respectively.}
		\label{Fig_2}
	\end{center}
\end{figure}

In Fig.~\ref{Fig_1} we have compared the exact result for $\left| \sum_{n=1}^\infty \frac{1}{E_n^2} - 1  \right| = \frac{\kappa^4}{64\pi^2}$, with the approximate sum rule obtained calculating the eigenvalues numerically with the Rayleigh-Ritz method, using the states with $1 \leq l \leq l_{max}$ and $|m|\leq l$, respectively with $l_{max}=30$ (dotted curve), $l_{max}=60$ (dashed curve) and $l_{max} = 90$ (dot-dashed curve).

The numerical sum rule is calculated using the lowest $N$ numerical eigenvalues ($N=320$, $1240$ and $2760$, respectively) and completing the series using the asymptotic behavior predicted by Weyl's law, $E_n^{(Weyl)} \approx n$, for $n \rightarrow \infty$:
\begin{eqnarray}
\sum_{n=1}^\infty \frac{1}{E_n^2}  \approx  \sum_{n=1}^{N} \frac{1}{\left(E_n^{(RR)}\right)^2} + \sum_{n=N+1}^\infty \frac{1}{\left(E_n^{(Weyl)}\right)^2} 
\end{eqnarray}

A similar result for the case of the sum rule of order three is displayed in Fig.~\ref{Fig_2}.

\section{Conclusions}
\label{conclusions}

We have used the method of Ref.~\cite{Amore14} to derive general integral formulas for the sums of 
inverse powers of the eigenvalues of the Laplacian on a heterogeneous sphere with arbitrary density. 
Due to the presence of a zero mode, i.e. of a mode with vanishing eigenvalue, the spectral sum rules
need to be "renormalized", by taking out the singular contribution of the fundamental mode: this 
is achieved by performing an infinitesimal shift $\gamma$ on the Laplacian, thus rendering all the eigenvalues
finite and then subtracting the contributions stemming from the lowest eigenvalue, for a finite infinitesimal
shift (calculated using perturbation theory). The resulting sum rule is now analytical at $\gamma=0$ and
it corresponds to the sum over the non-vanishing eigenvalues.

We have applied our general formulas to a non--trivial problem, corresponding to the variable density 
$\Sigma(\theta,\phi) = 1 +\frac{1}{2} \sqrt{\frac{3}{\pi}} \kappa \cos\theta$, with $|\kappa | < 2 \sqrt{\frac{\pi }{3}}$,
obtaining the exact expressions for the sum rules of order two and three as functions of $\kappa$.
These results have been verified numerically using the Rayleigh-Ritz method to calculate numerically the eigenvalues.

\section*{Acknowledgements}
The research of P.A. was supported by the Sistema Nacional de Investigadores (M\'exico).


\appendix

\begin{appendices}

\section{Corrections to the lowest eigenvalue}
\label{appA}

We report in the following the explicit expression for the corrections to the lowest eigenvalue of an heterogenous sphere calculated
using perturbation theory:
\begin{subequations}
\begin{align}
E_0^{(0)} &= 0 \nonumber \\
E_0^{(1)} &= \frac{4\pi}{\int \Sigma(\Omega) d\Omega } \nonumber\\
E_0^{(2)} &= -(4\pi)^2  \frac{\int \Sigma(\Omega) G^{(0)}(\Omega,\Omega') \Sigma(\Omega')  d\Omega d\Omega'} {\left(\int \Sigma(\Omega) d\Omega\right)^3 } \nonumber\\
E_0^{(3)} &= (4\pi)^2  \frac{\int \Sigma(\Omega) G^{(1)}(\Omega,\Omega') \Sigma(\Omega')  d\Omega d\Omega'}{\left(\int \Sigma(\Omega) d\Omega\right)^3 } \nonumber \\
&- (4\pi)^3  \frac{\int \Sigma(\Omega) G^{(0)}(\Omega,\Omega') \Sigma(\Omega') G^{(0)}(\Omega',\Omega'') \Sigma(\Omega'')  d\Omega d\Omega' d\Omega''} {\left(\int \Sigma(\Omega) d\Omega\right)^4 } \nonumber  \\
&+ 2 (4\pi)^3  \frac{\left(\int \Sigma(\Omega) G^{(0)}(\Omega,\Omega') \Sigma(\Omega')  d\Omega d\Omega' \right)^2} {\left(\int \Sigma(\Omega) d\Omega\right)^5 } \nonumber  \\
E_0^{(4)} &= -(4\pi)^2  \frac{\int \Sigma(\Omega) G^{(2)}(\Omega,\Omega') \Sigma(\Omega')  d\Omega d\Omega'} {\left(\int \Sigma(\Omega) d\Omega\right)^3} \nonumber \\
&+ 2 (4\pi)^3  \frac{\int \Sigma(\Omega) G^{(1)}(\Omega,\Omega') \Sigma(\Omega') G^{(0)}(\Omega',\Omega'') \Sigma(\Omega'') d\Omega d\Omega' d\Omega''} {\left(\int \Sigma(\Omega) d\Omega\right)^4 } \nonumber  \\
&-  \frac{(4\pi)^3}{\left(\int \Sigma(\Omega) d\Omega\right)^5}  \left[
\int \Sigma(\Omega) G^{(1)}(\Omega,\Omega') \Sigma(\Omega')  d\Omega d\Omega'  \int \Sigma(\Omega) G^{(0)}(\Omega,\Omega') \Sigma(\Omega')  d\Omega d\Omega'  \right. \nonumber \\ 
&+ \left. 4\pi
\int \Sigma(\Omega) G^{(0)}(\Omega,\Omega') \Sigma(\Omega') G^{(0)}(\Omega',\Omega'') \Sigma(\Omega'')  G^{(0)}(\Omega'',\Omega''') \Sigma(\Omega''')  d\Omega d\Omega'  d\Omega'' d\Omega''' \right]  \nonumber  \\
&+ 5 (4\pi)^4  \frac{\left(\int \Sigma(\Omega) G^{(0)}(\Omega,\Omega') \Sigma(\Omega')  d\Omega d\Omega' \right) \  \left( \int \Sigma(\Omega) G^{(0)}(\Omega,\Omega') \Sigma(\Omega') G^{(0)}(\Omega',\Omega'') \Sigma(\Omega'')  d\Omega d\Omega' d\Omega'' \right)} {\left(\int \Sigma(\Omega) d\Omega\right)^6 } \nonumber \\
&- 5 (4\pi)^4  \frac{\left(\int \Sigma(\Omega) G^{(0)}(\Omega,\Omega') \Sigma(\Omega')  d\Omega d\Omega' \right)^3} {\left(\int \Sigma(\Omega) d\Omega\right)^7 } \nonumber  
\end{align}
\end{subequations}


\section{Some integrals}
\label{appB}

We consider the density of the general form
\begin{equation}
\Sigma(\Omega) = 1 + \sum_{l=1}^\infty \sum_{m=-l}^l c_{lm} Y_{lm}(\Omega)
\end{equation}
where $c_{lm}$ are arbitrary coefficients such that $\Sigma(\Omega) > 0$ over the sphere.

The total mass is simply given by
\begin{equation}
\int \Sigma(\Omega) d\Omega = 4 \pi \nonumber
\end{equation}

Let us define the integrals:
\begin{subequations}
	\begin{align}
	\mathcal{I}_1^{(q)} &\equiv \int \Sigma(\Omega) G^{(q)}(\Omega,\Omega') \Sigma(\Omega')  d\Omega d\Omega' \nonumber \\
	\mathcal{I}_2^{(q,p)} &\equiv \int \Sigma(\Omega) G^{(q)}(\Omega,\Omega') \Sigma(\Omega')  G^{(p)}(\Omega,\Omega') \Sigma(\Omega'') d\Omega d\Omega' d\Omega'' \nonumber \\
	\mathcal{I}_3^{(q,p,r)} &\equiv \int \Sigma(\Omega) G^{(q)}(\Omega,\Omega') \Sigma(\Omega')  G^{(p)}(\Omega',\Omega'') \Sigma(\Omega'') G^{(r)}(\Omega'',\Omega''') \Sigma(\Omega''') d\Omega d\Omega' d\Omega''  d\Omega'''\nonumber \\
	\mathcal{J}_1^{(q, p)} &\equiv \int \Sigma(\Omega) G^{(q)}(\Omega,\Omega') \Sigma(\Omega') G^{(p)}(\Omega',\Omega) d\Omega d\Omega' \nonumber  \\
	\mathcal{J}_2^{(q, p,r)} &\equiv \int \Sigma(\Omega) G^{(q)}(\Omega,\Omega') \Sigma(\Omega') G^{(p)}(\Omega',\Omega'') \Sigma(\Omega'') G^{(r)}(\Omega'',\Omega) d\Omega d\Omega' d\Omega'' \nonumber  
	\end{align}
\end{subequations}

We have
\begin{subequations}
	\begin{align}
	\mathcal{I}_1^{(q)} &\equiv \int \Sigma(\Omega) G^{(q)}(\Omega,\Omega') \Sigma(\Omega')  d\Omega d\Omega' \nonumber \\
	&= \sum'_{l,m} \int \frac{Y_{lm}(\Omega) Y_{lm}^\star(\Omega')}{(l (l+1))^{q+1}} 
	\left(1 + \sum'_{l_1,m_1} c_{{l_1} m_1} Y_{{l_1} m_1}(\Omega)\right) \nonumber \\
	&\cdot \left(1 + \sum'_{l_2,m_2} c_{{l_2} m_2} Y_{{l_2} m_2}(\Omega)\right) d\Omega d\Omega' \nonumber \\
	&= \sum'_{l,m} \frac{|c_{lm}|^2}{(l (l+1))^{q+1}} 
	\end{align}
\end{subequations}
where we have defined $\sum'_{l,m} f_{lm} \equiv \sum_{l=1}^\infty \sum_{m=-l}^l f_{lm}$.

Similarly we can calculate the remaining integrals:
\begin{subequations}
	\begin{align}
	\mathcal{I}_2^{(q,p)} &= \sum'_{l,m}  \sum'_{l',m'} \int \frac{Y_{lm}(\Omega) Y_{lm}^\star(\Omega') Y_{l'm'}(\Omega') 
		Y_{l'm'}^\star(\Omega'')}{(l (l+1))^{q+1}  (l' (l'+1))^{p+1}} \nonumber \\
	& \left(1 + \sum'_{l_1,m_1}  c^\star_{{l_1} m_1} Y^\star_{{l_1} m_1}(\Omega)\right)  
	  \left(1 + \sum'_{l_2,m_2} c_{{l_2} m_2} Y_{{l_2} m_2}(\Omega')\right) \nonumber \\
	& \left(1 + \sum'_{l_3,m_3} c_{{l_3} m_3} Y_{{l_3} m_3}(\Omega'')\right)
	d\Omega d\Omega' d\Omega'' \nonumber \\
	&= \sum'_{l,m}  \frac{|c_{lm}|^2}{(l (l+1))^{p+q+2}} \nonumber \\
	&+ \sum'_{l,m}  \sum'_{l',m'}  \sum'_{l_2,m_2} 
	\frac{c_{lm}^\star c_{l'm'} c_{l_2 m_2}}{(l (l+1))^{q+1} (l' (l'+1))^{p+1}}
	W_{l,m,l',m',l_2,m_2} 
	\end{align}
\end{subequations}
where
\begin{subequations}
	\begin{align}
	W_{l_1,m_1,l_2,m_2,l_3,m_3} \equiv& \int Y_{l_1,m_1}^\star(\Omega) Y_{l_2,m_2}(\Omega) Y_{l_3,m_3}(\Omega) d\Omega \nonumber \\
	=& (-1)^{m_1} \sqrt{\frac{(2 l_1+1)(2 l_2+1)(2 l_3+1)}{4\pi}}  \ 
	\left( \begin{array}{ccc}
	l_1 & l_2 & l_3 \\
	0   &  0  &  0  \\
	\end{array}\right)\nonumber \\
	\cdot&
	\left( \begin{array}{ccc}
	l_1 & l_2 & l_3 \\
	-m_1   &  m_2  &  m_3  \\
	\end{array}\right) \nonumber
	\end{align}
\end{subequations}
and
\begin{subequations}
	\begin{align}
	\mathcal{I}_3^{(q,p,r)} &= \sum'_{l,m} \sum'_{l',m'} \sum'_{l'',m''} \int \frac{Y_{lm}(\Omega) Y_{lm}^\star(\Omega') Y_{l'm'}(\Omega') 
		Y_{l'm'}^\star(\Omega'')  Y_{l''m''}(\Omega'') Y_{l''m''}^\star(\Omega''')}{(l (l+1))^{q+1}  (l' (l'+1))^{p+1} (l'' (l''+1))^{r+1}} \nonumber \\
	& \left(1 + \sum'_{l_1,m_1}  c^\star_{{l_1} m_1} Y^\star_{{l_1} m_1}(\Omega)\right)  
	\left(1 + \sum'_{l_2,m_2} c_{{l_2} m_2} Y_{{l_2} m_2}(\Omega')\right) \nonumber \\
	& \left(1 + \sum'_{l_3,m_3} c_{{l_3} m_3} Y_{{l_3} m_3}(\Omega'')\right)
	 \left(1 + \sum'_{l_4,m_4} c_{{l_4} m_4} Y_{{l_4} m_4}(\Omega''')\right) d\Omega d\Omega' d\Omega'' d\Omega''' \nonumber \\
	&= \sum'_{l,m} \frac{|c_{lm}|^2}{(l (l+1))^{p+q+r+3}} \nonumber \\
	&+ \sum'_{l,m} \sum'_{l',m'} \sum'_{l_1,m_1} 
	c_{lm}^\star c_{l'm'} c_{l_1 m_1} W_{l,m,l',m',l_1,m_1} \nonumber \\
	& \cdot \left[
	\frac{1}{(l (l+1))^{q+1} (l' (l'+1))^{p+r+2}}
  	+ \frac{1}{(l (l+1))^{q+p+2} (l' (l'+1))^{r+1}}	\right] \nonumber \\
  	& + \sum'_{lm} \sum'_{l'm'}  \sum'_{l_2 m_2} \sum'_{l_3 m_3} \frac{c^\star_{l m} c_{l_2, m_2} c_{l_3, m_3} c_{l'' m''}}{(l (l+1))^{q+1} (l' (l'+1))^{p+1} (l'' (l''+1))^{r+1}}  W_{l,m,l',m',l_2,m_2} W_{l',m',l,m,l_3,m_3}
	\end{align}
\end{subequations}

Similarly we have
\begin{subequations}
	\begin{align}
	\mathcal{J}_1^{(q, p)} 
	&= \sum'_{lm} \sum'_{l'm'} \int \frac{Y_{lm}(\Omega) Y_{lm}^\star(\Omega') Y_{l'm'}(\Omega') Y_{l'm'}^\star(\Omega)}{(l (l+1))^{q+1} (l' (l'+1))^{p+1}}  \nonumber \\
	&\cdot \left(1 + \sum'_{l_1 m_1 } c_{{l_1} m_1} Y_{{l_1} m_1}(\Omega')\right) 
	\left(1 + \sum'_{l_2 m_2 } c^\star_{{l_2} m_2} Y^\star_{{l_2} m_2}(\Omega)\right) d\Omega d\Omega' 
	\nonumber \\
	&=  \sum'_{lm}  \frac{1}{(l (l+1))^{p+q+2}} 
	+ 2 \sum'_{lm}  \sum'_{l_1 m_1} \frac{c_{l_1,m_1}}{(l (l+1))^{p+q+2}} W_{l,m,l,m,l_1,m_1} \nonumber \\
	&+ \sum'_{lm} \sum'_{l' m'} \sum'_{l_1 m_1} \sum'_{l_2 m_2} 
	\frac{c^\star_{l_2,m_2} c_{l_1,m_1}}{(l (l+1))^{q+1} (l' (l'+1))^{p+1}} \nonumber \\
	&\cdot W_{l,m,l',m',l_1,m_1} W^\star_{l,m,l',m',l_2,m_2} 
	\end{align}
\end{subequations}
and
\begin{subequations}
	\begin{align}
	\mathcal{J}_2^{(q, p,r)} &= \sum'_{l,m} \sum'_{l',m'} \sum'_{l'',m''} \int \frac{Y_{lm}(\Omega) Y_{lm}^\star(\Omega') Y_{l'm'}(\Omega') 
		Y_{l'm'}^\star(\Omega'')  Y_{l''m''}(\Omega'') Y_{l''m''}^\star(\Omega)}{(l (l+1))^{q+1}  (l' (l'+1))^{p+1} (l'' (l''+1))^{r+1}} \nonumber \\
	& \left(1 + \sum'_{l_1,m_1}  c_{{l_1} m_1} Y_{{l_1} m_1}(\Omega)\right)  
	\left(1 + \sum'_{l_2,m_2} c_{{l_2} m_2} Y_{{l_2} m_2}(\Omega')\right) \nonumber \\
	& \left(1 + \sum'_{l_3,m_3} c_{{l_3} m_3} Y_{{l_3} m_3}(\Omega'')\right) d\Omega d\Omega' d\Omega''  \nonumber  \\
	& = \sum'_{lm} \frac{1}{(l (l+1))^{p+q+r+3}} 
	+ 3 \sum'_{lm} \ \sum'_{l_1 m_1} \frac{c_{l_1 m_1}}{(l (l+1))^{p+q+r+3}} 
	W_{l,m,l,m,l_1,m_1} \nonumber \\
	&+ 3 \sum'_{lm} \sum'_{l'm'}  \sum'_{l_1 m_1} \sum'_{l_2 m_2} 
	\frac{c_{l_1, m_1} c_{l_2,m_2}}{(l (l+1))^{q+1} (l' (l'+1))^{p+r+2}}
	W_{l',m',l,m,l_1,m_1} W_{l,m,l',m',l_2,m_2} 	
	\nonumber \\
	& + \sum'_{lm} \sum'_{l'm'} \sum'_{l''m''} \sum'_{l_1 m_1} \sum'_{l_2 m_2} \sum'_{l_3 m_3}  
	\frac{c_{{l_1} m_1} c_{{l_2} m_2} c_{{l_3} m_3}  }{(l (l+1))^{q+1} (l' (l'+1))^{p+1} (l'' (l''+1))^{r+1}} \nonumber \\
	& \cdot W_{l'',m'',l,m,l_1,m_1} W_{l,m,l',m',l_2,m_2} W_{l',m',l'',m'',l_3,m_3} 
	\end{align}
\end{subequations}

\end{appendices}


\begin{thebibliography}{Bibliography}	
\bibitem{Itzykson86} Itzykson, C., P. Moussa, and J. M. Luck, "Sum rules for quantum billiards." Journal of Physics {\bf A} 19 (1986): L111-L115.
\bibitem{Berry86} M.V. Berry, "Spectral zeta functions for Aharonov-Bohm quantum billiards", Journal of Physics {\bf A} 19 (1986): 2281-2296
\bibitem{Steiner87} Steiner, Frank. "Spectral Sum Rules for the Circular Aharonov‐Bohm Quantum Billiard." Fortschritte der Physik/Progress of Physics 35.1 (1987): 87-114.
\bibitem{Steiner85}	Steiner, F. "Magic sum rules for confinement potentials." Physics Letters B 159.4-6 (1985): 397-402.
\bibitem{Kvitsinsky96} Kvitsinsky, Andrei A. "Zeta functions of nearly circular domains." Journal of Physics A: Mathematical and General 29.19 (1996): 6379.
\bibitem{Dittmar02} Dittmar, Bodo. "Sums of reciprocal eigenvalues of the Laplacian." Mathematische Nachrichten 237.1 (2002): 45-61.
\bibitem{Dittmar11} B. Dittmar and M. Hantke, Annales UMCS, Mathematica, 65(2), 29-44 (2011)
\bibitem{Dostanic11} M.R. Dostani\'c, "Regularized trace of the inverse of the Dirichlet Laplacian." Communications on Pure and Applied Mathematics 64, 1148-1164  (2011)
\bibitem{Amore13A} Amore, Paolo. "Exact sum rules for inhomogeneous strings." Annals of Physics 338 (2013): 341-360.
\bibitem{Amore13B} Amore, Paolo. "Exact sum rules for inhomogeneous drums." Annals of Physics 336 (2013): 223-244.
\bibitem{Amore14} Amore, Paolo. "Exact sum rules for inhomogeneous systems containing a zero mode." Annals of Physics 349 (2014): 253-267.
\bibitem{Amore18} Amore, Paolo, "Exact sum rules for quantum billiards of arbitrary shape", Annals of Physics 388 (2018): 12-24.
\bibitem{Amore10} Amore, Paolo, "Spectroscopy of drums and quantum billiards: Perturbative and nonperturbative results", J. Math. Phys. {\bf 51}, 052105 (2010); doi: http://dx.doi.org/10.1063/1.3364792
\bibitem{Szmytkowski06} Szmytkowski, Radosław. "Closed form of the generalized Green’s function for the Helmholtz operator on the two-dimensional unit sphere." Journal of mathematical physics 47.6 (2006): 063506.
\bibitem{Freeden80} Freeden, William, "On integral formulas of the (unit) sphere and their application to numerical computation of integrals." Computing 25.2 (1980): 131-146.
\bibitem{Englis98} Englis, Miroslav, and Jaak Peetre. "Green's functions for powers of the invariant Laplacian." Canadian Journal of Mathematics 50.1 (1998): 
40-73.
\bibitem{Gustavsson01} Gustavsson, Jan. "Some sums of Legendre and Jacobi polynomials." Mathematica Bohemica 126.1 (2001): 141-149.



\end{thebibliography}
\end{document}